**Shell elasticity and viscosity of lipid-coated microbubbles are significantly altered in mediums of different ionic strength**

Amin Jafari Sojahrood, [abc], C. Yang [ab], C. Counil[d], P. Nittachayarn[d], D.E. Goertz[c], A.A Exner[d] and M.C. Kolios[abc]

a. Department of Physics, Toronto Metropolitan University, Toronto, Ontario, Canada.

b. Institute for Biomedical Engineering, Science and Technology (iBEST), A Partnership Between Toronto Metropolitan University and St. Michael's Hospital, 209 Victoria Street, Toronto, Ontario, Canada

c. Department of Medical Biophysics, University of Toronto, Ontario, Canada

d Department of Radiology, Case Western University, Cleveland, Ohio, USA

**ABSTRACT**. Correct measurement of the shell properties of coated microbubbles (MBs) is essential to understanding and optimizing their response to ultrasound (US) exposure parameters in diagnostic and therapeutic ultrasound. MBs are surrounded by blood; however, the influence of the surrounding medium charges on the MB properties is poorly understood. This study aims to measure the medium charge interactions with MB shells by measuring the frequency-dependent attenuation of the same size MBs in mediums of varying charge density. In-house lipid-coated MBs with $C_3F_8$ gas core were made. The MBs were isolated to a mean size of 2.35μm using differential centrifugation. MBs were diluted to ~$8*10^5$ MBs/mL in distilled water (DW), and two different concentrations of phosphate buffered saline solution (PBS-1x and PBS-10x). The frequency-dependent attenuation of the MBs solutions was measured using an aligned pair of PVDF transducers with a center frequency of 10MHz and 100% bandwidth in the linear oscillation regime (7 kPa pressure amplitude). The MB shell properties were estimated by fitting the linear equation to experiments. Using a pendant drop tensiometer, the surface tension of mm-size drops was measured inside DW, PBS-1x and PBS-10x. The surface tension at the gas/solution interface was estimated by fitting the Young-Laplace equation from the recorded images. The frequency of the peak attenuation changes at different salinity levels was 13, 7.5 and 6.25MHz in DW, PBS-1x and PBS-10x, respectively. The attenuation peak increased by ~140% with increasing ion density. MBs' estimated shell elasticity decreased by 64% between DW and PBS-1x and 36% between PBS-1x and PBS-10x. Reduction in the shell stiffness is in qualitative agreement with the drop surface tension measurements. The shell viscosity was reduced by ~40% between DW and PBS-1x and 42% between PBS-1x and PBS-10x. The reduction in the stiffness and viscosity is possibly due to the formation of a densely charged layer around the shell, further reducing the effective surface tension on the MBs. This effect may be utilized in enhancing the MB oscillation amplitude and a better understanding of the interaction of the magnetic fields with MBs during MRI-guided applications.

1. Introduction

Acoustically excited bubbles have fundamental roles in various applications, ranging from cleaning [1] and food processing [2] to sonochemistry [3] and medical ultrasound [4-7]. In medicine, microbubbles (MBs) in the 1-10μm range are encapsulated by a stabilizing shell that can be polymer, protein or phospholipid based [4]. They are used in medical imaging as contrast agents to visualize the blood flow [5] and have potential applications in targeted drug delivery [6], blood-brain barrier opening with human clinical trials underway [7], and at higher acoustic pressures, they have potential applications for thermal [8] and mechanical ablation of tumors [9].

Due to the fundamental role of bubbles in many applications, the dynamics of the bubbles have been studied extensively, and it is well-known that the bubble oscillator is nonlinear and complex [10-20]. Control and optimization of the bubble behavior in applications are, therefore, challenging and require detailed knowledge about the response of the bubbles to the acoustic exposure parameters.

The bubble response to ultrasonic excitation depends on the bubble size, the surrounding medium and the inner gas composition [4, 10-20]. In medical applications, the shell adds an extra parameter that affects the MBs dynamics, mainly by increasing the resonance frequency and damping of the MBs [4, 10]. In the case of lipid-coated MBs, the buckling and rupture of the shell can substantially alter the MB response to the ultrasound field [4, 15-20]. Due to the strong influence of the shell and the surrounding medium on the MB response, many studies aimed to develop theoretical models, and several investigations attempted to characterize the MB shell properties [21-25]. Despite the considerable number of studies investigating the influence of MB shell characteristics on the MB response to ultrasound, several aspects of the problem have not been thoroughly investigated.

One important but under-investigated parameter is the influence of the surface charges on the MB response to ultrasound. The impact of the accumulation of ions on the MB response is not well known. The surface charge effects on uncoated bubbles were studied in early 1900s [26-27] and later revisited by [28] in 1992 to theoretically explain the presence of stable microscopic gas bubbles in a pure liquid with no contaminants. This problem has been recently revisited by [29-30] to find an explanation for the long stability of bulk nano-sized bubbles (NBs) seemingly in defiance of the "Laplace pressure catastrophe". The stability of NBs was linked to the accumulation of charges at the NB surface, reducing the surface tension and, consequently, the Laplace pressure [29]. Despite strong potential bubble-charge interactions, however, only a few numerical studies investigated the influence of the charges on the nonlinear bubble dynamics. Although these studies [31-33] are limited to parametric explorations, they demonstrated substantial effects on bubble dynamics.

The influence of selective charge adsorption has not been investigated using carefully controlled experiments. This problem becomes more significant in medical ultrasound, where MBs are inside blood which is dense with ions. Importantly, the encapsulating lipid shell is charged, potentially enhancing the surface ion adsorption that can alter the effective MB surface tension and, therefore, MB response to ultrasound. However, to our best knowledge, no study has investigated this important effect on coated MBs. This study aims to shed light on this topic. In this paper, we investigate the medium charge interactions with the MB shell by measuring the frequency-dependent attenuation of lipid-coated MBs in mediums of varying charge density, and by fitting the appropriate models, the change in shell properties will be investigated.

2. Materials and Methods

2.1 MB formulation

Ultra-stable lipid shell MBs were prepared as described in detail in [34] and supplementary material (S1). 1 mL of prepared lipid solution was allocated in 3 mL glass vials which were capped with rubber septa and aluminum seals using a vial crimper. The room air in the vials was removed and exchanged for octofluropropane gas (99.9% purity $C_3F_8$, Electronic Fluorocarbons). Samples were then activated using a VialMix (Bristol-Myers Squibb Medical Imaging Inc.) for 45 s. The resultant MB solution is polydisperse. The polydisperse MB solution was then suspended in distilled water (DW) in a 25mL syringe and size-isolated to a batch of mean size of 2.35µm using a 5-stage differential centrifugation [35] protocol.

2.2 **Attenuation measurements**

Size-isolated MBs were diluted to ~$8*10^5$MBs/mL in distilled water (DW), PBS-1x and PBS-10x for attenuation measurements. The attenuation measurements were performed using a pair of 100% bandwidth PVDF transducers (10MHz center frequency, 2.2cm active diameter, Precision Acoustics). The frequency-dependent attenuation of the MB solution was calculated by subtracting the measured spectra of the MB suspension from that of the pure water using equation 1 [36]:

$$\alpha(f) = \frac{20}{d} \log_{10} \frac{|FFT_{pure\ water}(f)|}{|FFT_{MB}(f)|}$$

Eq. (1)

where α is the attenuation, $d$ is the sample holder thickness (acoustic path length of 0.5cm). The experimental setup is shown in Fig. S1, accompanied by more information about the setup details. The measurements were performed in the linear regime of the oscillations with a negative peak pressure of 7kPa. Experiments were repeated a minimum of 3 times.

2.3 Surface tension measurements

To assess the effect of different salinity levels on the lipid monolayer surface and its potential effects on the MB's effective surface tension, pendant drop tensiometry was applied as described in detail in [37] and supplementary material S3. The surface tension at the $C_3F_8$–water interface was measured for the lipid solution in DW, PBS-1x and PBS-10x. After the formation, the images were acquired, and the image dimensions were fitted using the Young–Laplace equation.

2.4 Linear model

The shell stiffness and viscosity were estimated by fitting the linear model to the measured attenuation curves using Eq. 1 [4, 22,36]:

$$\alpha(f) = \frac{10}{\ln(10)} \sum_i n(R(i))\sigma_s(R(i),f) \frac{\delta_{to}(R(i),f)}{\delta_{rad}(R(i),f)}$$

Eq. 1

where $\alpha$ is the attenuation, $f$ is frequency, $R(i)$ and $n$ are the radius and number density of the *ith* MB in the size bin, $\sigma_s(R(i),f)$, $\delta_{rad}(R(i),f)$ and $\delta_{tot}(R(i),f)$ are the frequency-dependent scattering cross section, radiation damping constant and total damping constant of MB with radius $R(i)$, respectively [4,22,36]. Summation is performed over all the size bins.

3. Results and Discussion

Figure 1a shows a representative size distribution of the MBs used in experiments. Figure 1b shows the frequency-dependent attenuation of the MBs in DW, PBS-1x and PBS-10x at 7kPa excitation amplitude. With increasing salinity, the peak frequency of the attenuation curve decreases, the attenuation curves' width decreases, and the peak attenuation increases, indicative of potential reductions in the shell stiffness and viscosity. The frequency of the attenuation peak is 13, 7.5 and 6.25 MHz in DW, PBS-1x and PBS-10x, respectively. The peak attenuation is ~8.6, 15.5 and 20.8 dB/cm in DW, PBS-1x and PBS-10x, respectively. The optical MB size measurements (not shown here) did not show any differences between the MBs samples; thus, the changes in the resonance frequency and attenuation peak may be attributed to the changes in the effective shell properties.

Results of the surface tension measurements are shown in Fig. 2a. The drop surface tension decreases with increasing salinity (66.5 mN/m in DW, 59.5mN/m in PBS-1x and 56.5 mN/m in PBS-10x.) Fig. 2b shows the MBs estimated shell stiffness ($S_p$) decreases by 64% between DW (7 N/m) and

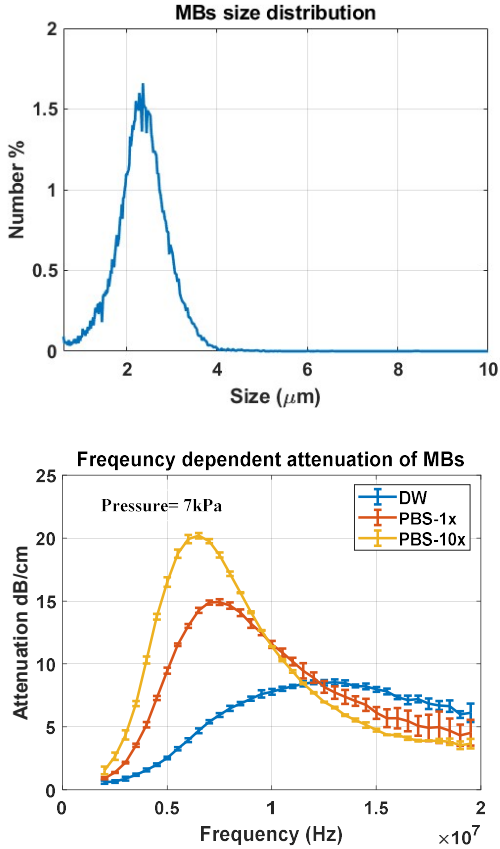

Figure 1: a) The size distribution of the MBs, b) the frequency dependent attenuation of the MBs in DW, PBS-1x and PBS-10x at pressure amplitude of 7kPa (n=3, error bars represent standard deviation).

PBS-1x (2.5N/m) and 36% between PBS-1x and PBS-10x (1.6 N/m). Reduction in the shell stiffness is in qualitative agreement with drop surface tension measurements despite a large difference between the drop size (mm range) and MBs (μm range). The shell viscosity ($S_f$ in Fig. 2c) is reduced by ~40% between DW (1 μkg/s) and PBS-1x (0.6 μkg/s) and by ~42% between PBS-1x and PBS-10x (0.35 μkg/s). The reduction in the stiffness and viscosity is possibly due to the formation of a densely charged layer around the shell, further reducing the effective surface tension on the MBs. By increasing the salinity of the medium, the density of the charged ions at the MB surface increases, thus further reducing the effective surface tension on the MB. This manifests as a reduction in the resonance frequency (~ frequency of the attenuation peak) and an increase in the MB oscillation amplitude (the attenuation peak). In the case of uncoated bubbles, the reduction in the surface tension by charge effects may be expressed as $\sigma - Q^2/(16\pi\varepsilon R^3)$ [26,27,31] where $\sigma$ is the water surface tension, and $Q$ is the total surface charge, $R$ is the MB radius and $\varepsilon$ is the electrical permittivity of the surrounding medium. The higher the ion adsorption at the interface, the higher the surface tension reduction, which is qualitatively seen in the estimated effective shell stiffness. It is hypothesized [29-30] that the zeta potential of NBs is greater than of MBs, and thus NBs strongly stabilize themselves by accumulating surface charges preventing the Laplace pressure catastrophe [29-30]. The experimental observations of the surface tension reduction in this paper support this hypothesis as a general point in principle. Future studies using NBs (~400-800nm) and higher sonication frequencies are needed to shed light on this topic.

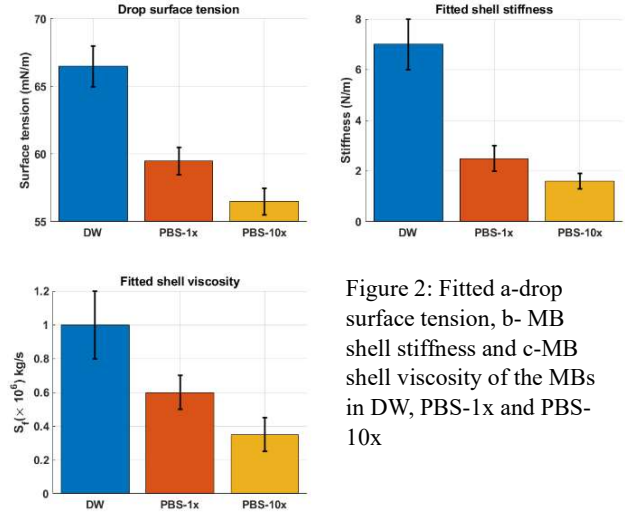

Figure 2: Fitted a-drop surface tension, b- MB shell stiffness and c-MB shell viscosity of the MBs in DW, PBS-1x and PBS-10x

One of the open problems in the characterization of the shell properties of the MBs is the dependence of the stiffness and viscosity on the MB radius, despite having the same shell material. It has been suggested that the lipid shell may exhibit shear-thinning and stiffness softening [25, 38,39]. Here we also show that the surface charges influence the shell stiffness and viscosity. The zeta potential is size dependent [29,30]. More charges may accumulate for smaller MBs; therefore, the smaller shell stiffness and viscosity of the smaller MBs may have resulted from the surface charge effects. Future studies using different MB sizes are needed to confirm this hypothesis and quantify size-dependent charge effects.

Recently it has been shown that the MB oscillations undergo additional damping under the influence of strong Magnetic fields (1-4T) during MRI-guided blood-brain barrier opening [40]. The Magnetic pressure on an oscillating MB can be expressed as $\Upsilon B^2 R\dot{R}/\rho$ [41, 42] where $\Upsilon$, B, $\dot{R}$ and $\rho$ are the medium conductivity, magnetic field strength, MB wall velocity and medium density, respectively. However, the observed strong damping cannot be expressed using the magnetic pressure expression considering the minimal blood conductivity value (20 mS/cm), and requires unphysical magnetic field strengths of the order of $10^4$T [41,42]. However, this expression is obtained in the absence of surface charges. Conductivity is directly proportional to the surface charge density. A stronger conductivity is expected for the liquid layer adjacent to lipid shell MBs in the blood due to the accumulation of ions. Future studies of MB shell stiffness in blood plasma and subsequent MB oscillations under magnetic fields of varying strengths are needed to better understand and quantify this effect.

4. Conclusions

The influence of the medium salinity on the dynamics of lipid-coated MBs was studied by measuring the frequency-dependent attenuation of the MBs in DW, PBS-1x and PBS-10x for the first time. Results showed a strong stiffness softening and shell viscosity reduction with increasing the medium ion density (e.g. 140% increase in the attenuation peak). This effect may have consequences in increasing the MB stability. It may also enhance the MB oscillations in blood at a specific frequency (by designing an appropriate shell and potentially increasing the contrast-to-tissue ratio). It may also provide a better understanding of the stability of NBs and MB oscillations in blood and the interaction of magnetic fields with MBs during MRI-guided applications.

ACKNOWLEDGEMENTS

This work is supported in part by the National Institute of Health (R01EB025741 and R01EB028144). Dr. Sojahrood is also supported by Canada NSERC Postdoctoral Fellowship.

pressure dependence of sound speed and attenuation in bubbly media: Experimental observations, a theoretical model and numerical calculations. Ultrasonics Sonochemistry, 95, p.106319.

## Supplementary information

### 1-Lipid solution preparation

A lipid combination consisting of 60.1 mg of 1,2-dibehenoyl-sn-glycero-3-phosphocholine (DBPC C22; Avanti Polar Lipids), 10 mg of 1,2 dipalmitoyl-sn-glycero-3-phosphate (DPPA; Avanti Polar Lipids), 20 mg of 1,2-dipalmitoyl-sn-glycero-3-phosphoethanolamine (DPPE; Avanti Polar Lipids), and 10 mg of 1,2-distearoyl-sn-glycero-3-phosphoethanolamine-N-[methoxy(polyethyleneglycol)-2000] (DSPE-mPEG2000; Avanti Polar Lipids) was dissolved in 1mL of propylene glycol and was heated in a water bath at 80 °C until full lipids dissolved completely. 1 mL of glycerol and 8 mL of phosphate-buffered saline (PBS-1X) were then heated to 80 °C and added to the lipid solution. After mixing, the resulting solution was sonicated in a bath sonicator (Branson Sonicator CPX2800H) for 15 min at room temperature.

### 2- Experimental setup

The experimental setup (Fig. S1) consists of two transducers facing each other and operating in transmit and receive mode. A broadband signal (0-50MHz) was generated using a pulser receiver (DPR300-JSR Ultrasonics). The pulse was then sent to one of the transducers (PVDF transducer with 10MHz center frequency – 100%BW-Precision Acoustics). The second transducer received the generated acoustic pressure wave after propagation through the sample hold with 5mm thickness and 7.5 cm diameter. The distance between the two transducers is 15cm, with the sample holder in the middle. The signals were acquired in pure DW water and cases of diluted MBs in DW, PBS-1x and PBS-10x.

### 3-Drop surface tension measurements

An air-tight syringe with a 1 mm diameter J-shaped needle was filled with octafluoropropane ($C_3F_8$) gas. In the cuvette, 1 mL of the lipid solution was added. The lipid solution was diluted with 5 mL of DW, PBS-1x or PBS-10x. A gas bubble of 4 ± 1 mL of the $C_3F_8$ gas was generated at the tip of the J-shaped needle. For each sample, images (Fig. S2) of the rising drops were acquired every 5 seconds using a KSV CAM 200 Optical Contact Angle Meter equipped with a charge-coupled device camera. All measurements have been done at least in triplicate. The equipment software then analyzed images to get the drop dimension. Fitting the Young–Laplace equation on the recorded images provides the surface tension estimate at the gas/water interface. To validate the calibration, the surface tension of pure water (Milli-Q water with a resistivity of ~18 MΩ·cm) was also measured, expecting a surface tension of 72.1 mN/m.

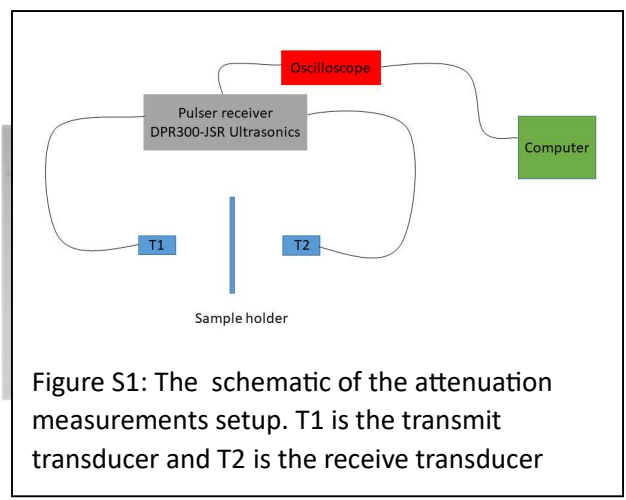

Figure S1: The schematic of the attenuation measurements setup. T1 is the transmit transducer and T2 is the receive transducer

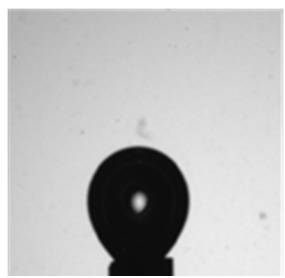

Figure S2: A ample image of the drop solution in DW.

## 4-Identification of the linear oscillation regime

Figure. S3 shows the attenuation spectra of the MBs suspension in DW at various pressure amplitudes between 3.5kPa to 120kPa. The attenuation spectra are in the linear regime for pressures at and below 15kPa. Increasing the pressure from 3.5kPa to 15kPa does not change the attenuation curve; however, for pressures at and above 45kPa, the frequency of the peak attenuation shifts toward lower amplitudes while the attenuation peak increases. This is known as the pressure-dependent attenuation [4,24-25,43], and the pressure-dependent resonance frequency of the MBs is the physical mechanism behind it. By confining the MB sizes, we could smoothly detect the pressure at which the transition between the linear and nonlinear regime occurs. The experimental data at 7kPa thus were used when fitting to the linear model.

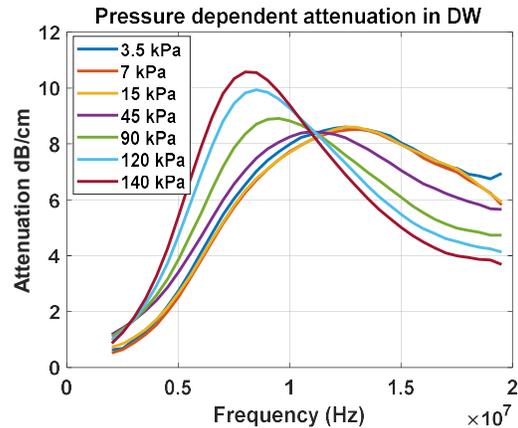

Figure S3: Frequency-dependent attenuation of the MBs solution in DW at various pressures